\documentclass{aa}  
\usepackage{textcomp} 
\usepackage{graphicx}
\usepackage{txfonts}

\usepackage[dvipsnames]{xcolor}
\usepackage{float}
\usepackage{longtable,booktabs}
\usepackage{amsmath}	
\usepackage{multirow}
\usepackage{url}
\usepackage{ae,aecompl}
\usepackage[flushleft]{threeparttable}
\usepackage{wasysym} 
\usepackage{scalerel} 
\usepackage{verbatim} 
\usepackage{listings}
\usepackage{txfonts}
\usepackage{natbib,twoopt}
\usepackage{xspace}
\usepackage{balance}
\usepackage{upgreek}
\usepackage{array}
\usepackage{soul}
\usepackage{outlines}
\usepackage{makecell}
\usepackage{siunitx}
\usepackage{tabularx}
\usepackage{lipsum}
\usepackage[normalem]{ulem}
\usepackage[switch]{lineno}  

\usepackage[colorlinks=true, allcolors=blue]{hyperref}

\definecolor{grey}{rgb}{0.4,0.5,0.6}
\definecolor{darkgreen}{rgb}{0.0,0.45,0.0}
\definecolor{darkorange}{rgb}{0.9,0.2,0.0}
\definecolor{ballblue}{rgb}{0.13, 0.67, 0.8}
\definecolor{blush}{rgb}{0.87, 0.36, 0.51}

\begin{document} 

\title{Halo Occupation Distribution estimation performance for LSST data}

\author{P. Cataldi \inst{1}, V. Cristiani \inst{2,3,4}, F. Rodriguez \inst{2,3,4}, A. Taverna \inst{2,5,6},  M.C. Artale \inst{7}, B. Levine \inst{8} \& the LSST Dark Energy Science Collaboration}

\institute{
        Instituto de Astronom\'{\i}a y F\'{\i}sica del Espacio, CONICET-UBA, Casilla de Correos 67, Suc. 28, 1428, Buenos Aires, Argentina
        \email{pcataldi@iafe.uba.ar}
    \and
        Instituto de Astronom\'ia Te\'orica y Experimental -  Observatorio Astron\'omico de C\'ordoba (IATE, UNC--CONICET),  C\'ordoba, Argentina
    \and
        Observatorio Astron\'{o}mico de C\'{o}rdoba, Universidad Nacional de C\'{o}rdoba, Laprida 854, X5000BGR, C\'{o}rdoba, Argentina
    \and
        Facultad de Matem\'atica, Astronom\'{\i}a y F\'{\i}sica, Universidad Nacional de C\'ordoba (FaMAF--UNC). Bvd. Medina Allende s/n, Ciudad Universitaria, X5000HUA, C\'ordoba, Argentina
    \and
        Instituto de Astronomia, Universidad Nacional Autonoma de Mexico, Apdo. Postal 106, Ensenada 22800, B.C., Mexico
    \and
        Center for Theoretical Physics, Polish Academy of Sciences, Al. Lotników 32/46, 02-668 Warsaw, Poland
    \and
        Universidad Andres Bello, Facultad de Ciencias Exactas, Departamento de Fisica y Astronomia, Instituto de Astrofisica, Fernandez Concha 700, Las Condes, Santiago RM, Chile
    \and
        Department of Physics and Astronomy, Stony Brook University, Stony Brook, NY 11794, USA
    }

\date{Received XX, XXXX; accepted XX, XXXX}

 
  \abstract
  {Upcoming imaging surveys, such as the Vera C. Rubin Observatory Legacy Survey of Space and Time (LSST), will enable high signal-to-noise measurements of galaxy clustering. The halo occupation distribution (HOD) is a widely used framework to describe the connection between galaxies and dark matter haloes, playing a key role in evaluating models of galaxy formation and constraining cosmological parameters. Consequently, developing robust methods for estimating this statistic is crucial to fully exploit data from current and future galaxy surveys.}
  {The main goal of this project is to extend a background subtraction method to estimate the HOD with more photometry-based information in preparation for the clustering analysis of the upcoming LSST data and to enable the study of the HOD with significantly improved statistical power. We evaluate the performance of the method using a mock galaxy redshift survey constructed from the \textsc{cosmoDC2} catalogue.}
  {We implement an extension of the background subtraction technique to utilize information from photometric galaxy surveys. To identify the centres of galaxy groups, we implement an iterative centroiding approach (Central Galaxy Finder). We evaluate the impact of each step in our pipeline, including group size estimation from luminosity and \textit{purity}, and \textit{completeness} on group identification, along with the influence of observational systematics such as the use of photometric redshifts and halo mass uncertainties.}
  {We demonstrate the validity of the proposed method using a mock galaxy catalogue, recovering the HOD from \textsc{cosmoDC2} over the absolute magnitude range $M_r = -20.0$ to $-17.0$ and halo masses up to $10^{15}\, \mathrm{M_\odot}$. We present key performance metrics to quantify the precision and reliability of the group finder and the resulting HOD measurements.}
  {}


   \maketitle
%

\section{Introduction}
\label{section:intro}

Understanding how galaxies trace the underlying dark matter distribution is a central goal of modern astrophysics. In the $\Lambda$ Cold Dark Matter ($\Lambda$CDM) paradigm, galaxies form within the gravitational potential wells of dark matter haloes where structure formation is hierarchical: smaller haloes collapse first and subsequently merge to build larger systems \citep{White1978}. However, the diversity and complexity of the astrophysical processes involved in galaxy formation and evolution pose significant challenges in determining exactly how galaxies occupy and populate dark matter haloes.

The link between galaxies and haloes is fundamental for several reasons. First, it provides a way to infer cosmological constraints. Galaxy surveys allow us to study the spatial distribution of haloes through measures such as galaxy clustering, and infer key cosmological parameters \citep{Bosch2003,Zheng2005}. Furthermore, it provides a framework to explain the physics behind galaxy formation, allowing the contribution of central and satellite galaxies to be studied separately \citep{Kravtsov2004,Cooray2005,White2007,Yang2008}. Observationally, central galaxies are identified as those located near the centres of galaxy groups or clusters, or as galaxies lacking comparably bright neighbours. Theoretically, they are expected to reside close to the minimum of the gravitational potential of their host dark matter haloes. Satellite galaxies correspond to the remaining group members.

Two commonly used statistical approaches to describe the galaxy--halo connection are the Conditional Luminosity Function (CLF) and the Halo Occupation Distribution (HOD). The CLF \citep[e.g.][]{Bosch2003,Yang2005,Bosch2007} describes the average number of galaxies with luminosity $L \pm dL/2$ that reside in haloes of mass $M_h$. The HOD \citep[e.g.][]{Jing1998,Peacock2000,Zheng2005,Tinker2005} defines the probability distribution $P(N|M_h)$, representing the likelihood that a halo of mass $M_h$ hosts $N$ galaxies with certain properties, typically distinguishing between central and satellite populations. While both approaches are closely related, the CLF is typically framed in terms of luminosity, whereas the HOD focuses on galaxy number counts.

The development of HOD models in the early 2000s \citep[e.g.][]{Seljak2000,Wechsler2001,Scoccimarro2001,Berlind2002} represented a major advance in linking galaxy clustering with halo statistics. These models have since been used to constrain both galaxy formation scenarios and cosmological parameters. Later studies have refined and applied the HOD formalism across a wide range of galaxy samples \citep[e.g.][]{Mandelbaum2006,Brown2008,Agustsson2006,More2009,Cacciato2009,Neistein2011,Avila-Reese2011,Leauthaud2012}, confirming its effectiveness in connecting observed galaxies to the underlying dark matter structure. The HOD of dark matter haloes typically follows a simple form: satellite occupation scales with halo mass via a power-law relation ($N \propto M_h ^{\alpha} $), and each halo typically hosts a central galaxy \citep{Kravtsov2004}.

A key challenge in studying satellite galaxies is the limited availability of spectroscopic data. Any satellite-counting exercise is complicated by the faintness of the satellites, which makes obtaining redshift information difficult. To address this, \citet{Liu2011} proposed a method based on photometric data to identify candidate satellites surrounding central galaxies, followed by statistical corrections for foreground and background contamination. This approach can be further improved using a background subtraction technique (hereafter BST), which is effective for analysing galaxy populations statistically \citep{Lares2011}. Several studies, from early works \citep{Holmberg1969,Phillipps1987} to more recent efforts \citep{Rodriguez2015,Duckworth2019}, have demonstrated the value of combining spectroscopic and photometric surveys to detect satellite galaxies at fainter magnitudes.

\citet{Rodriguez2015} validated the BST for estimating HODs using mock catalogues constructed from semi-analytic galaxy formation models within the Millennium simulation \citep{Springel2005}. They found a strong overall agreement between the HODs obtained via BST and those derived from direct galaxy counts in volume-limited samples across different absolute magnitude thresholds. Despite a slight overestimation at the faint end ($M_{\mathrm{lim}} \approx -16.0$), the method was shown to be reliable across the full magnitude range ($-21.5<M_r<-17.0$) \citep[see Fig.~1 in][]{Rodriguez2015}.

In this work, we seek to extend the BST for estimating the HOD using galaxy catalogues primarily based on photometric data. Unlike previous implementations, such as \citet{Rodriguez2015}, which relied on spectroscopically confirmed group positions and masses, our method requires a catalogue for estimating total group masses, but uses angular positions and apparent magnitudes for the rest. This approach reduces the reliance on spectroscopic observations, enabling the inclusion of much larger and deeper samples. In particular, it facilitates the study of more distant galaxy groups and broadens the accessible mass range of the resulting group samples.

Successfully adapting BST to photometric data would make it applicable to current and forthcoming wide-field surveys such as the Vera C. Rubin Observatory Legacy Survey of Space and Time \citep[LSST;][]{Abell2009, LSST2012}, Roman \citep{Spergel2015, Akeson2019}, and Euclid \citep{Laureijs2011}. These surveys will deliver unprecedented sky coverage, depth, and galaxy number densities—LSST alone will observe billions of galaxies out to $z \sim 3$ and identify around 100,000 galaxy clusters up to $z \sim 1$. A BST optimised for photometric data would therefore provide a timely and powerful tool to exploit these datasets, significantly enhancing the statistical power of HOD studies.

This paper is structured as follows. In  Sect.~\ref{section:data} we present the mock catalogue and the object sample used in this work. Sect.~\ref{section:method} and  Sect.~\ref{section:CGF} describe the background subtraction and central galaxy identification techniques. Sect.~\ref{section:results} outlines our testing of the method and details the methods applied for halo mass assignment, and photometric redshift estimation. Finally, we summarise our conclusions in Sect.~\ref{section:conclusions}.

\section{Data}
\label{section:data}

This work is based on the Data Challenge 2 (DC2) dataset \citep{Korytov2019, Kovacs2022} developed by the LSST Dark Energy Science Collaboration \citep{Abolfathi2021}, which covers 440~deg$^2$ and is designed for testing and developing LSST analysis pipelines. The core of DC2 is the \textsc{cosmoDC2} galaxy catalogue, built from the Outer Rim N-body simulation \citep{Heitmann2019}. Dark matter haloes were identified using a friends-of-friends (FoF) algorithm \citep{Davis-1985} (linking length of $b=0.168$), with two mass definitions provided: the total FoF mass $M _{\mathrm{FoF}}$ and the spherical overdensity (SO) mass $M_{\mathrm{200}}$. The SO of the haloes is defined as the mass within a sphere of radius ($r_{\mathrm{200}}$), whose mean internal density is 200 times the critical density of the Universe at redshift $z$.

Galaxies were populated using UniverseMachine \citep{Behroozi2019} and GalSampler \citep{Hearin2020}, with their physical and photometric properties generated through the Galacticus semi-analytic model \citep{Benson2010}. Spectral energy distributions (SED) were built from stellar population models using \textsc{fsps} \citep{Conroy2013}. The final catalogue includes approximately 2.26 billion galaxies, each with $\sim$550 properties—such as stellar mass, morphology, magnitudes in six LSST bands, redshift, and lensing quantities. The simulation assumes an idealised setting, free of dust extinction or stellar contamination. Although stellar masses are taken directly from the simulation and do not enter the BST—contributing only indirectly to the validation tests—luminosities are used to estimate the group radius, as described in Subsection~\ref{section:SS_2}. These luminosities are computed in the observer frame by blue-shifting the filter transmission functions according to the cosmological redshift (with peculiar velocities neglected), while $k$-corrections are interpolated to the lightcone redshifts to minimise discreteness effects in the observed colour space.

To account for photometric redshift uncertainties, we use the \textsc{cosmoDC2} photo-$z$ add-on catalogues produced by DESC: FlexZBoost  \citep{Izbicki2017} and BPZ \citep[Bayesian Photometric Redshifts;][]{Benitez2011}. FlexZBoost is based on an empirical machine learning method that estimates conditional redshift distributions $P(z|m)$ from photometry in the LSST bands. BPZ is a template-fitting approach that incorporates SED models and galaxy-type priors. FlexZBoost and BPZ have demonstrated high accuracy in previous studies \citep[e.g.;][]{Schmidt2020,Payerne2025} and are suited for statistical applications such as HOD estimation. 

\subsection{Group and galaxy catalogues}
\label{section:sample}

Galaxy groups can be identified via their member galaxies using the galaxy catalogues of \textsc{cosmoDC2}. Importantly, selecting groups directly from the \textsc{cosmoDC2} galaxy catalogue ensures that the resulting group catalogue remains unaffected by atmospheric or instrumental effects. The \textsc{cosmoDC2} catalogue includes information on galaxy group positions, redshifts, and halo masses. 

We construct a mock catalogue using synthetic galaxies extracted from a semi-analytic model of galaxy formation applied to the \textsc{cosmoDC2} simulation. We build this catalogue by collecting a comprehensive set of galaxy properties, including angular positions on the sky (\textsc{ra}, \textsc{dec}), halo id (\textsc{halo{\_}id}), magnitud in the r LSST bands (\textsc{mag{\_}r{\_}lsst}), stellar and halo masses (\textsc{stellar{\_}mass}, \textsc{halo{\_}mass}), and a flag indicating whether each galaxy is central (\textsc{is{\_}central}). We also include the \textit{true} redshift (\textsc{redshift}), the corresponding photometric redshift estimates from FlexZBoost and BPZ, (\textsc{photo{\_}z{\_}mean}), and the derived redshift quality metric (\textsc{photo{\_}z{\_}Odds}). From this full dataset, we select galaxies with a \textit{true} redshift $z < 0.4$ and apparent $r$-band magnitude $m_r < 21$, resulting in a sample of approximately $1.15 \times 10^6$ galaxies. The magnitude cut ensures high completeness, as $m_r < 21$ can be easily achieved with the LSST single-visit depth, while the redshift limit guarantees that we retain all potential satellite galaxies associated with the highest redshift volume-limited sample used in this work ($z_{\max} = 0.304$).

Constructing volume-limited samples is crucial to mitigate redshift-dependent incompleteness resulting from observational limits on apparent magnitude. At higher redshifts, fainter galaxies fall below the detection threshold, making it increasingly difficult to identify low-luminosity systems. This incompleteness can indirectly impact the halo mass range being probed, since lower-mass haloes tend to contain lower-mass, fainter galaxies, which are preferentially lost at higher redshift.

\begin{figure}
    \centering    
    \includegraphics[width=\columnwidth]{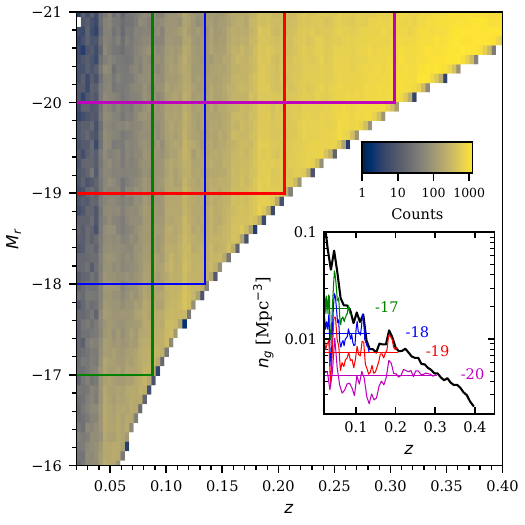}
    \caption{Construction of volume-limited galaxy samples with varying luminosity thresholds. The colour map shows the distribution of \textsc{cosmoDC2} galaxies as a function of \textit{true} redshift $z$ and $r$-band absolute magnitude $M_{r}$. The different coloured lines delineate the selection boundaries for the volume-limited samples. The \textit{inset panel} shows their mean galaxy number density $n_{\mathrm{g}}$ (colour lines) and total sample density (black line). The solid horizontal lines show the average number density for each sample (the values are given in Tab. \ref{tab:tab1}).}
    \label{fig:fig1}
\end{figure}

To ensure completeness, we define several absolute magnitude limits, $M_{\mathrm{r}} < -17$, $-18$, $-19$, and $-20$, each corresponding to a higher redshift cut and a brighter selection threshold. For each magnitude cut, we apply a redshift limit to guarantee that the brightest galaxy in each group remains observable across the entire volume. The selection criteria applied to the galaxy catalogue is illustrated in Fig.~\ref{fig:fig1}, while the resulting subsample properties---including the number of galaxies, number of central galaxies, and their number densities---are summarised in Tab~\ref{tab:tab1}.

\begin{table}
\caption{Galaxy samples used to test the method described in subsection~\ref{section:SS_1}. Group centres correspond to central galaxies with absolute magnitudes $M_r < -19.5$. Each row represents a different selection based on magnitude and completeness cuts. The mean galaxy number density  $n_{g}$ is in units of $10^{-3} \mathrm{Mpc ^{-3}}$.}
\label{tab:tab1}
\begin{center}
\small
\renewcommand{\arraystretch}{1.3}  
\begin{tabularx}{\columnwidth}{X|X|X|X|X}
\hline
 $M_{\mathrm{lim}}$ & $z_{\mathrm{max}}$ & $N_{\mathrm{gal}}$ & $N_{\mathrm{cen}}$ & $n_{g}$\\
\hline
$-17$  & $0.088$ &  $4.1 \cdot 10^{4}$ & $8.8 \cdot 10^{3}$ & 19.3\\
$-18$  & $0.135$ &  $8.5 \cdot 10^{4}$ & $2.9 \cdot 10^{4}$ & 11.2\\
$-19$  & $0.206$ &  $1.9 \cdot 10^{5}$ & $1.1 \cdot 10^{5}$ & 7.5\\
$-20$  & $0.304$ &  $3.5 \cdot 10^{5}$ & $3.2 \cdot 10^{5}$ & 4.6\\
\hline
\end{tabularx}
\end{center}
\end{table}

\begin{figure}
    \centering
    \includegraphics[width=\columnwidth]{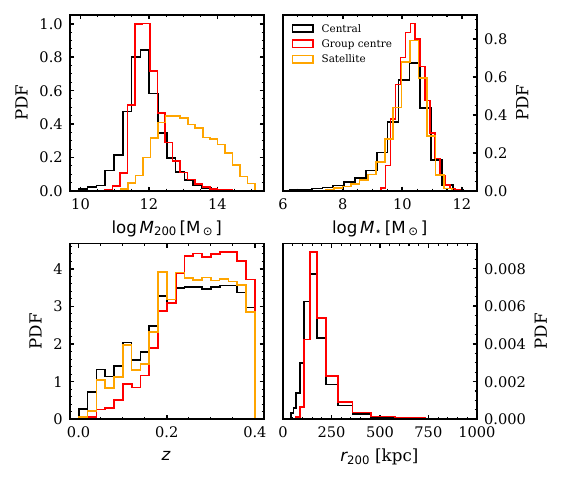}
    \caption{Normalized distributions for our galaxy catalogue, shown separately for centrals (black), group centres (red), and satellites (orange) galaxies. Panels show the distributions of halo mass $M_{\mathrm{200}}$ (top left), stellar mass $M_{\star}$ (top right), galaxy redshift $z$ (bottom left); and halo radius $r_{\mathrm{200}}$, derived from the halo mass (bottom right). The halo radius distribution includes only the central and group centre samples. All distributions are shown for the combined sample used in our analysis, which includes galaxies across the four absolute magnitude cuts.}    
    \label{fig:fig2}
\end{figure}

We restrict the luminosity of group members: they must contain at least one bright galaxy to be considered a group. We define a bright galaxy as one with an absolute $r$-band magnitude $M_r < -19.5$ \footnote{This particular choice is motivated by consistency with the methodology adopted in previous works \citep{Yang2008, Tempel2009, Rodriguez2020, Yang2021}}, following \citet{Rodriguez2020} and \citet{Zandivarez2024}. Thus, in our analysis, a group of galaxies is a gravitationally bound system that contains at least one bright (central) galaxy. The central galaxies in this magnitude-limited subset are used to define the group centres.

Central and satellite galaxies exhibit different distributions for their properties, as shown in Fig.~\ref{fig:fig2}. The black histogram represents all true central galaxies, while the red histogram shows the group centres defined above. Yellow lines indicate satellite galaxies, defined as those not labelled as centrals in the simulation.

As expected, the halo mass distribution for central galaxies presents a peak at a lower mass than that of satellites. This arises from the fact that low-mass haloes tend to have one (central) galaxy, while massive haloes are likely to host satellite galaxies. Both true centrals and our group centres show similar stellar mass peaks; however, the latter tend to be slightly more massive on average. This reflects the fact that their selection is based on brightness, which naturally favours galaxies with higher stellar and halo masses, excluding fainter, low-mass centrals. 

\begin{figure}
    \centering
   \includegraphics[width=\columnwidth]{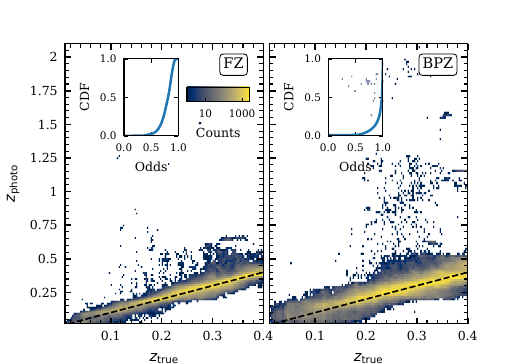}
    \caption{(Left) Comparison between central galaxies’ \textit{true} redshifts ($z_{\mathrm{true}}$) and photometric redshifts using FlexZBoost ($z_{\mathrm{photo}}$, mean value). (Right) same as left but for BPZ redshifts. The dashed line represents the one-to-one relation ($z_{\mathrm{photo}} = z_{\mathrm{true}}$). The \textit{inset panels} show the cumulative distribution function (CDF) of Odds parameters.}
    \label{fig:fig3}
\end{figure}

In this work, we investigate the impact of photometric redshifts on the performance of the BST at determining group centres. To that end, we use for group centres the photometric redshift add-on \textsc{cosmoDC2} catalogues derived from FlexZBoost and BPZ. Fig.~\ref{fig:fig3} presents a comparison between \textit{true} redshifts ($z_{true}$) and photometric redshifts ($z_{photo}$). The dashed lines represent the one-to-one relation, highlighting where photometric redshifts perfectly match the true values. Despite being unbiased, there is a considerable scatter between the photometric and \textit{true} redshifts, with FlexZBoost providing lower dispersion than BPZ. Nevertheless, FlexZBoost results should be considered optimistic due to the deep selection of the training complete sample up to $i<25$  (and thus unrealistic for real data), while the ``discreteness" in the colour-redshift of the modelled galaxies \citep{Korytov2019} negatively impacts the performance of BPZ. As pointed out in \citet{Payerne2025}: the actual LSST data will most likely fall between the pessimistic BPZ and optimistic FlexZBoost runs we analysed.

This uncertainty in redshift, and therefore in inferred galaxy positions, significantly complicates the recovery of galaxy clustering signals. Nonetheless, photometric surveys offer the advantage of much larger galaxy samples. Large errors in the photometric redshift are especially common for faint blue galaxies, whose colours can appear similar at both low and high redshifts, making them difficult to distinguish in photometric surveys. The quality-derived redshift \textsc{photo{\_}z{\_}Odds} parameter (hereafter, Odds) provides additional insight into the reliability of a photometric redshift estimate (see \textit{inset panels} of Fig.~\ref{fig:fig3}). It quantifies how peaked the redshift probability distribution function (PDF) is by integrating the PDF within a defined interval around its mode. A high Odds value indicates a narrow, well-constrained PDF, suggesting that the algorithm is confident in a specific redshift solution. In contrast, a low Odds value implies that the PDF is either broad or multi-modal, reflecting greater uncertainty or the presence of multiple plausible redshift solutions. In the context of this work, we test the performance of photometric redshift estimates and, where appropriate, apply an Odds-based selection to our galaxy catalogue to obtain more constrained subsamples. We note, however, that while an Odds cut can significantly remove a good deal of the outliers, it does not eliminate them entirely and may also remove a substantial number of well-estimated galaxies with point estimates near the true redshifts.

\section{Background subtraction technique}
\label{section:method}

The BST has been validated through comparisons with other HOD estimation methods based on observational data. BST can yield results consistent with those obtained from spectroscopically identified galaxy groups, while enabling the study of satellite populations at fainter absolute magnitudes than previously accessible \citep[see Fig.~5 in][]{Rodriguez2015}. The method has been successfully applied in various contexts, including the investigation of the environmental dependence of kinematically misaligned galaxies using MaNGA data \citep{Duckworth2019}, and the study of the mass–richness relation using galaxy catalogues from the Sloan Digital Sky Survey \citep{Gonzalez2016}. 

\begin{figure}
\centering    
\includegraphics[width=\columnwidth]{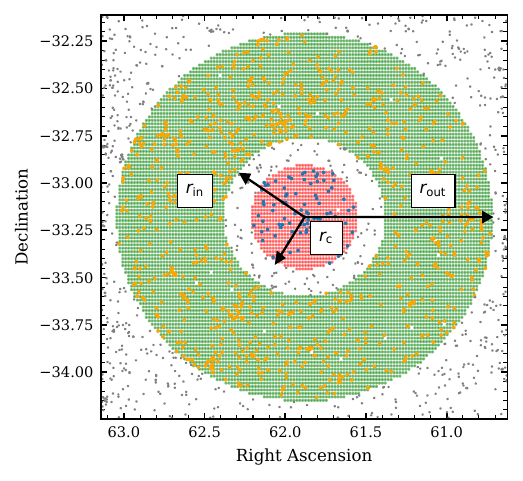}
\caption{Illustration of the BST for a single group. The red-shaded area and blue dots indicate the region and galaxies located within the characteristic radius of the host halo ($r < r_{\mathrm{c}}$). The green-shaded area and yellow dots represent the region and galaxies located in the outer annulus ($r_{\mathrm{in}} < r < r_{\mathrm{out}}$). Gray dots correspond to galaxies that lie outside the region considered by the method.}
\label{fig:fig4}
\end{figure}

The background subtraction procedure assumes that the large-scale distribution of galaxies is isotropic and homogeneous, and that galaxy groups represent local overdensities. This technique allows estimation of the number of galaxies that appear to belong to a group in projection, but are in fact located in front of or behind it (commonly referred to as background galaxies, although some may lie in the foreground), using statistical analysis. Given the sky area and depth of LSST-like photometric catalogues, by combining spectroscopic surveys for central galaxies with photometric galaxy catalogues, we can estimate the HOD over a wider magnitude range and improve the overall statistics. The primary advantage of this approach is its ability to determine the HOD of satellite galaxies to fainter magnitudes than previously studied. 

In Fig.~\ref{fig:fig4}, we show the application of BST to a single group extracted from the \textsc{cosmoDC2} simulation. The figure illustrates the circular aperture used to select candidate members, and the surrounding annuli employed to estimate the local background density. To apply the BST, we follow the steps detailed below:

\begin{itemize}

    \item Redshift assumption: Given that only photometric information is available, we assume that all galaxies are located at the same redshift as the central galaxy of the group, $z_{\mathrm{gr}}$. In this work, we first adopt the true redshift for the group centres, and subsequently extend the analysis to the case where photometric redshifts are used. The absolute magnitude is then computed as:
    \begin{equation}
        M = m - 25 - 5 \log (d_L(z_{gr}))
    \label{eq:eq1}
    \end{equation}
    where $d_L(z_{gr})$ is the luminosity distance in Mpc at $z_{gr}$. 

    \item Galaxy counting: Define $n_i$ as the number of galaxies with absolute magnitude $M_r < M_{\mathrm{lim}}$ within a circle centred in each group with a radius determined by some projected characteristic radius of the halo ($r_{c}$), which is often taken to be $r_{\mathrm{200}}$. The value of $M_{\mathrm{lim}}$ sets the upper limit for the range of absolute magnitudes where the HOD will be estimated.

    \item Background estimation: A statistical method is necessary, as the background contribution to $n_i$ is challenging to determine directly. Given the hierarchical nature of large-scale structure, it is known that a given overdensity is always embedded in a larger density mode, which may cause the matter density nearby to be greater or less than the cosmic mean. To estimate the number of background galaxies ($n_0$), count those that satisfy the same magnitude criterion in an annulus ($r_{in}<r<r_{out}$) around each group, rather than using a global average.

    \item Background subtraction: To discount the contribution of background galaxies, which cannot be distinguished from the group galaxies, we need to determine their number density in the region surrounding the group. We compute the corrected number of galaxies $N$ in each group using:
    \begin{equation}
        N = n_i - \frac{n_0}{A_0}A_i
    \label{eq:eq2}
    \end{equation}
    where $n_0$ is the galaxy count in the annulus, and $A_i$ and $A_0$ are the projected areas of the group circle and background annulus, respectively.

    \item Sample selection: Since the method has no restrictions on the choice of groups, we should be careful when selecting the samples used in each measurement. Specifically, low-mass groups cannot be detected at high redshifts because they typically contain faint galaxies, while high-mass groups can be observed across the entire redshift range. This is addressed by setting a redshift limit $z_{\mathrm{max}}$ for each mass bin, such that the brightest galaxy in each group is detectable.

    \item HOD estimation: Bin the groups by halo mass and compute the HOD as the average number of galaxies per group: $\left< N | M_h \right>$, considering only those groups for which galaxies with $M_{\mathrm{lim}}$, located at $z_{\mathrm{gr}}$, would have an apparent magnitude brighter than the limiting apparent magnitude of the photometric catalogue.
\end{itemize}

\subsection{Extension to photometric data}

Traditionally, implementing BST requires detailed information for each group, including halo mass, redshift, angular position, and a characteristic radius, alongside a galaxy catalogue containing angular positions and magnitudes.

We adapt this framework to operate solely on photometric catalogues, removing the reliance on spectroscopically confirmed group properties. To achieve this, we incorporate a complementary method to identify potential central galaxies directly from photometric data. This approach is referred to as the \textit{Central Galaxy Finder} (CGF) and is outlined in Sect.~\ref{section:CGF}. We also leverage available photometric redshift estimates to statistically constrain galaxy membership and estimate the HOD.

Our approach is more challenging due to the inherent uncertainty in photometric redshifts, but it enables the application of the method to the full depth and area of upcoming wide-field photometric surveys. Moreover, this implementation lays the groundwork for a wide range of future studies on the galaxy--halo connection, including halo assembly bias, galactic conformity, and secondary clustering dependence on properties such as colour and environment density.
\section{\textit{Central Galaxy Finder} (CGF)}
\label{section:CGF}

In this work, we extend the BST algorithm to enable its application to galaxies with only photometric information. To make this extension feasible, it is first necessary to identify potential galaxy groups—or at least their central galaxies—using photometric data alone, since the BST method requires accurate group positions. 

\begin{figure}
 \centering
 \includegraphics[width=\columnwidth]{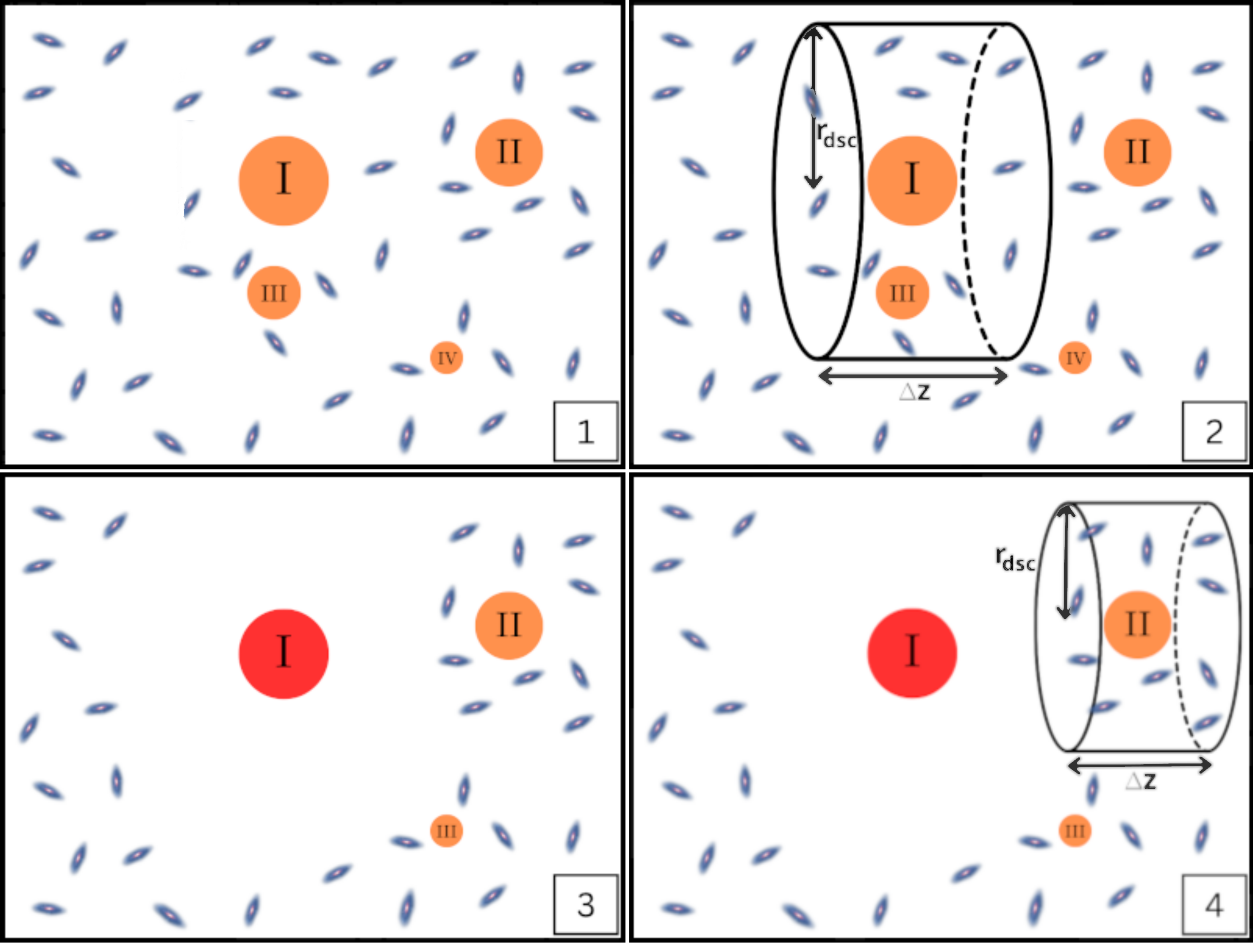}
    \caption{Scheme of the CGF method applied to a galaxy sample. Top left: selection of candidate group centres, ranked by brightness (orange dots). Top right: definition of a discard region centred on the brightest galaxy (I), determined by its luminosity. Bottom left: removal of all galaxies within the discard region from the candidate list, except for galaxy (I). Bottom right: continue with the next remaining brightest galaxy. In all panels, the line of sight is along the direction perpendicular to the plane of the plot (i.e. viewed from the side of the plot).}
    \label{fig:fig5}
\end{figure}

In this context, we introduce a method to identify central galaxy candidates based on their brightness relative to the local galaxy density. This step provides the necessary input catalogue of group positions and properties to which the BST can be applied. Alternatively, other group finder algorithms designed to be used in large photometric surveys can also be used in conjunction with the BST, such as redMaPPer \citep{Rykoff2014}, CFSFDP \citep{Zou2021}, and WaZP \citep{Aguena2021}. Below, we outline the main steps employed in our \textit{Central Galaxy Finder} (CGF, see Fig.~\ref{fig:fig5} for a schematic illustration.)

Step 1: Initial Selection. We begin by selecting all galaxies with an absolute magnitude in the $r$-band of $M_{r} < -19.5$. These galaxies are considered potential group candidates and are ordered in decreasing luminosity. Step~1 is illustrated in panel~1 of Fig.~\ref{fig:fig5}, showing the full sample of luminous galaxies.

Step 2: Discard-radius estimation. Starting with the brightest galaxy in the list, we estimate its discard group radius using the relation $r_{\mathrm{dsc}} = f_r \cdot r_c $, where $f_r$ is a free scaling parameter and $r_c$ is the previously mentioned characteristic radius of the halo. The radius $r_{\mathrm{dsc}}$ defines the projected search region for potential group members and is referred to as the \textit{discard radius}. This projected volume for the brightest galaxy is shown in panel~2 of Fig.~\ref{fig:fig5}.

Step 3: Member Selection. We select all galaxies that lie within a projected distance $r < r_{\mathrm{dsc}}$ and have a redshift difference $\Delta z < f_z$, where $\Delta z = |z_{\mathrm{can}} - z_{\mathrm{gal}}|$. Here, $z_{\mathrm{can}}$ is the redshift of the brightest galaxy, and $z_{\mathrm{gal}}$ is the redshift of the other candidates. The parameter $f_z$ is another free variable of the method. All selected galaxies are marked as members of the group, and they are then removed from the catalogue for subsequent steps. Panel~3 of Fig.~\ref{fig:fig5} shows the catalogue after discarding the identified group members.

Step 4: Iteration. We move to the next brightest galaxy that has not been previously discarded, and repeat Steps 2 and 3. Panel~4 in Fig.~\ref{fig:fig5} illustrates this step, where the selection moves to the next remaining bright galaxy. This iterative process continues until the entire catalogue has been processed.
        
For reliable HOD estimation, high central galaxy \textit{purity} across the full halo mass range is essential—and also the most challenging to achieve. \textit{Purity} is defined as the percentage of selected candidates that are truly central galaxies, and \textit{completeness} is defined as the fraction of true centrals that are correctly identified. 

\begin{figure}
    \centering    
    \includegraphics[width=\columnwidth]{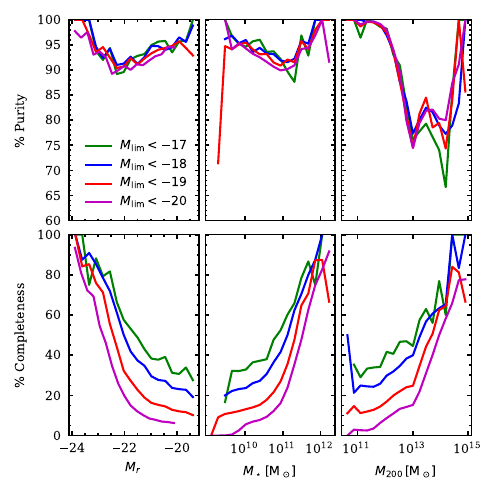}
    \caption{(Top panels) \textit{Purity} of the resulting catalogue as a function of absolute magnitude (left), stellar mass (middle), and halo mass (right). The galaxies were selected from the CGF method with a fixed $f_r =2.5$ and a redshift bin width of $f_z = 0.25$. (Bottom panels) Corresponding \textit{completeness} shown as a function of the same variables. Different coloured lines represent the median values for the volume-limited samples.}
    \label{fig:fig6}
\end{figure}

In Fig.~\ref{fig:fig6}, we present the resulting \textit{purity} and \textit{completeness} of the identified central galaxy candidates for a fixed set of parameters: $(f_r,f_z) = (2.5,0.25)$, shown as a function of absolute magnitude (left panel), stellar mass (middle), and halo mass (right), with different coloured lines indicating median values under various magnitude cuts representing different observational selections. The choices of this fixed set of parameters improve the \textit{purity} of true centrals (above $\sim$70\% over halo mass) by excluding interlopers over a larger projected and redshift volume. Although this configuration results in lower \textit{completeness}, this is less concerning for future photometric surveys due to their large statistical samples. 

The characteristic U-shaped trend in purity across $M_r$, $M_\star$, and most prominently $M_{200}$, arises naturally from the way the CGF operates. Because the CGF initially selects the brightest galaxies in the catalogue, these early candidates are, in the vast majority of cases, \textit{true} centrals. They are also less likely to be removed in the iterative discard steps of the algorithm. This leads to both high \textit{completeness} and high \textit{purity} at the bright/massive end. As the CGF continues through the candidate list, progressively fainter galaxies are examined. At this stage, completeness decreases because more galaxies are discarded during iteration, including some genuine centrals. Purity also declines because the removal of a \textit{true} central can result in an “orphan” group, for which the CGF assigns the brightest remaining galaxy in the local region—often a satellite. This effect drives the drop in purity at intermediate magnitudes, stellar masses, and halo masses.

At the faint/low-mass end, the completeness is indeed lower (these galaxies are considered last in the CGF iteration cycle), but most of the corresponding haloes host only one or two members. In such cases, the CGF classification is typically correct, which produces the recovery in \textit{purity} at the faint end. Finally, the more pronounced U-shape observed when \textit{purity} is plotted against $M_{200}$ reflects the fact that the CGF selection is based on brightness (and therefore indirectly on $M_{\star}$), so \textit{purity} is expected to be systematically lower and to show stronger variations across halo mass bins.

To further understand the trade-offs involved, Fig.~\ref{fig:fig_apx1} (Appendix~\ref{sec:apx_CGF}) presents the \textit{purity} and \textit{completeness} of the CGF as a function of halo mass ($M_{\mathrm{200}}$) in the range $10^{12}$–$10^{15} \,\mathrm{M_{\odot}}$, focusing on this high-mass regime where achieving high \textit{purity} for central galaxies is particularly challenging. We find a clear tension between \textit{purity} and \textit{completeness} across different combinations of $f_r$ and $f_z$: high values of both parameters increase the exclusion volume in each iteration of the CGF, resulting in high \textit{purity} but reduced \textit{completeness}. Conversely, lower values of $f_r$ and $f_z$ improve \textit{completeness} by retaining more candidates, but at the cost of lower \textit{purity} due to increased contamination. Given that future surveys such as LSST will provide excellent statistics through large galaxy samples, we opt for a parameter set that prioritises higher \textit{purity}, ensuring more accurate central galaxy identification and thus more reliable HOD recovery. 

The CGF naturally favours isolated, bright galaxies, leading to high purity when selections are based on stellar mass or magnitude. Every group-finding algorithm introduces selection biases arising from the trade-off between purity and completeness. Our CGF-derived central-galaxy catalogue remains broadly consistent with the true centrals from the simulation (see Fig.~\ref{fig:fig_apx2}). For comparison, \citet{Yang2021} reported $\gtrsim 80 \%$ completeness in member detection but only $\gtrsim 70 \%$ of groups reaching high purity, and their method is reliable only for haloes above $10^{12.5} \,\mathrm{M}_{\odot}$. Likewise, \textsc{redMaPPer} achieves an almost unbiased cluster catalogue but operates at even higher masses ($M \gtrsim 10^{14} \,\mathrm{M}_{\odot}$), well above the regime relevant for our HOD analysis, where lower-mass systems are more prone to misclassification and mass overestimation.

\section{Results}
\label{section:results}

In the following sections, we present a step-by-step evaluation of our method. In Subsection~\ref{section:SS_1}, we validate the BST using simulation-based quantities such as \textit{true} redshifts, angular positions, halo masses, and central galaxy identifiers to define group centres. Subsection~\ref{section:SS_2} explores an alternative approach to estimating group sizes by substituting halo size with luminosity derived from absolute \textit{r}-band magnitudes. In Subsection~\ref{section:SS_3}, we investigate the impact of FlexZBoost and BPZ photometric redshifts on the inference of the HOD. Subsection~\ref{section:SS_4} assesses the impact of uncertainties in halo mass estimation. Finally, in Subsection~\ref{section:SS_5}, we evaluate the performance of our method on the CGF outputs applied to photometric data. 

\subsection{Evaluation of the Method}
\label{section:SS_1}

\begin{figure*}
    \includegraphics[width=\textwidth]{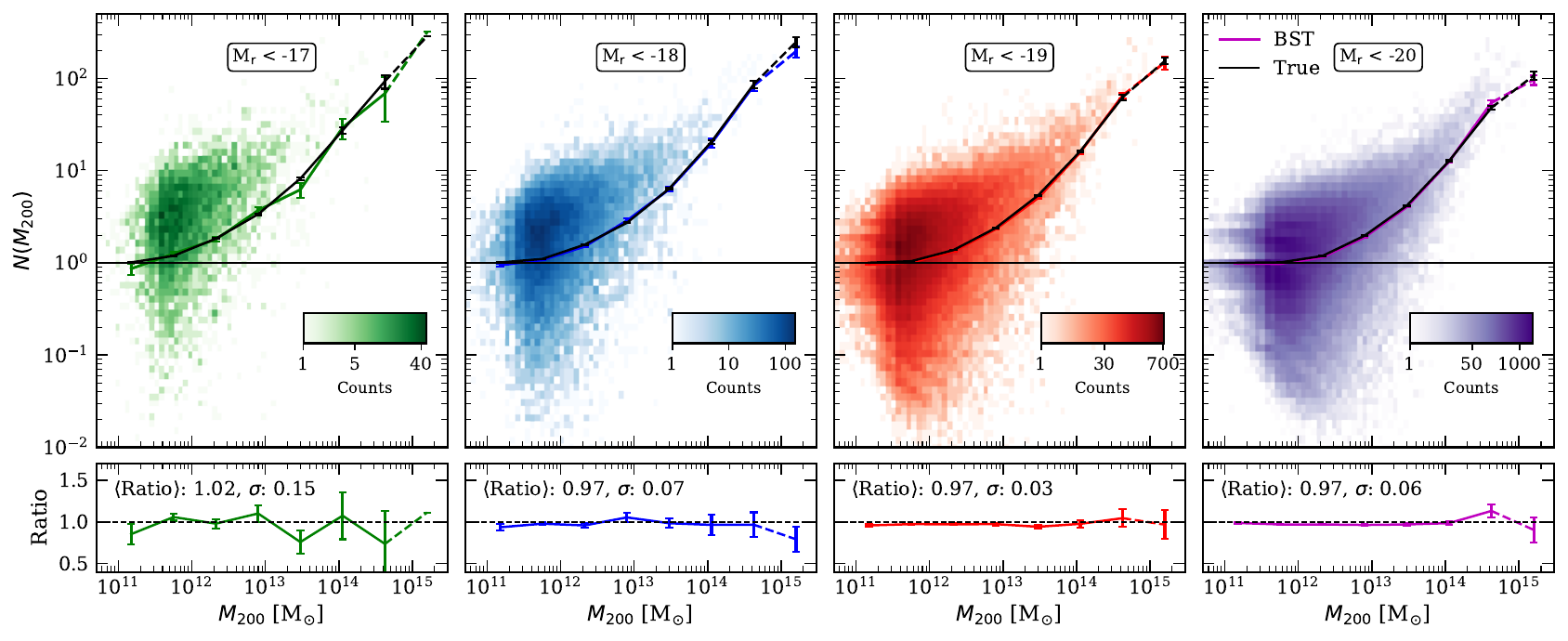}
    \caption{Comparison between BST results (colour lines) and the \textit{true} HOD obtained by direct counts in volume-limited samples (black lines). Each panel shows results for $M_{\mathrm{\lim}}=-17,-18,-19$ and -20. The colour maps represent the occupation number obtained for each volume-limited sample according to the BST, while the coloured lines indicate the median values of the corresponding distributions. Dashed lines correspond to the highest halo-mass bins ($M_\mathrm{200} > 5 \times  10^{14} \, \mathrm{M_\odot}$), which are affected by low-number statistics. For a better quantification of the differences, the lower panels show the ratio between the BST-recovered and true HOD, where we show the median and dispersion values. All uncertainties were calculated by the standard jackknife procedure.}
    \label{fig:fig7}
\end{figure*}

This section aims to evaluate the performance of BST. To this end, we use the mock catalogue introduced earlier, which provides full knowledge of galaxy populations within dark matter haloes, enabling a direct comparison between the true group properties and those recovered by our method.

As a first test, we apply the technique under idealised conditions: using the true angular positions, redshifts, and halo masses of central galaxies to define group locations, along with the true angular positions of satellite galaxies. This setup isolates the intrinsic accuracy of the method prior to introducing observational uncertainties.

We adopt a characteristic radius of $r_{\mathrm{c}}=r_{200}$, as it serves as a good upper limit for group sizes, defined as the maximum Cartesian distance between the central galaxy and its most distant satellite. The \textsc{cosmoDC2} intra-halo distribution of satellites follows a truncated NFW profile \citep{Korytov2019}, so it is expected that the satellites population drops sharply after $r>r_{200}$. 

We impose a maximum radius of $r_{\mathrm{c}} = 1$ Mpc to avoid excessively broad annuli. This ensures that the computed local galaxy density is not diluted by including regions representative of the cosmic mean density, which would otherwise lead to an overestimation of group membership in massive clusters. Only a small fraction of groups exceed this upper limit: about $0.2 \%$ for $M_{\mathrm{lim}}=-17$, and even lower fractions for the brighter samples. Therefore, this cut-off has a minimal impact on the overall catalogue.

To compute the background contribution as described in Sect.~\ref{section:method}, we define an annular region of  $1.5\,r_{\mathrm{c}}<r<3.5\,r_{\mathrm{c}}$. We tested different combinations of inner and outer radii of the annular region, specifically $r_{\mathrm{in}} = \{1.5, 2.0, 2.5\}$ and $r_{\mathrm{out}} = \{3.5, 4.0, 4.5, 5.0\}$, and found that the BST results remain stable across all choices, as long as the annular region avoids including actual group members, but is not so large that it samples environments unrepresentative of the local density in the vicinity of the group.

Figure~\ref{fig:fig7} shows the HOD across a broad range of halo masses. The true HOD is computed from the known galaxy group memberships in the mock catalogue by binning in halo mass and averaging the number of galaxies per bin. The black lines represent the median true HOD. The background colour maps display the two-dimensional distribution of galaxy counts recovered using the BST, while the coloured lines show the median HOD for different galaxy samples defined by varying magnitude limits. Uncertainties were estimated using a random jackknife sample, in which the BST output is randomly divided into 50 equal subsets and the HOD is recomputed while omitting each subset in turn. The lower panels show the ratio between the BST and the true HOD, for which we compute the median and dispersion as a quantitative measure of the BST performance.

We find that the agreement between the BST-recovered and true HOD generally improves as the magnitude limit becomes brighter (i.e., lower $M_{\mathrm{lim}}$), which increases the effective redshift range (up to $z < 0.3$) and enhances sample completeness. This improvement is also due to the increased sample size in these brightness bins (see Tab.~\ref{tab:tab1}), leading to lower statistical scatter. We note that the median and dispersion of the ratio between the BST-recovered and true HOD are typically below unity; however, in all cases, they remain consistent with unity within the quoted uncertainties. Across the halo mass range, we observe both underestimation and overestimation of the HOD, indicating no systematic bias of the BST-recovered HOD. The offset between the median and the mode of the BST  distribution arises from the way the BST computes the number of galaxies in each group through Eq.~\ref{eq:eq2}, which can yield $N(M_{\mathrm{200}}) \le 0$ for some groups, particularly at low-density environments. In Fig.~\ref{fig:fig7} we display the results in logarithmic scale, excluding these non-positive values. Nevertheless, all $N(M_{\mathrm{200}})$ are included when computing the median relation shown in the HOD analysis.

The highest halo mass bins ($M_\mathrm{200} > 5 \times 10^{14} \,\mathrm{M_{\odot}}$), indicated with a dashed line, suffer from poor statistics due to the limited number of massive groups in the \textsc{cosmoDC2} simulation. Consequently, the median values in this regime should be interpreted with caution. We include these results for completeness, noting their statistical limitations, and because the recovered HOD still shows reasonable agreement. This issue is expected to be significantly alleviated with future datasets from wide-area surveys such as LSST, which will cover $\sim$18,000\,deg$^2$ and provide a vastly larger sample of both galaxies and massive clusters

Interestingly, the mock catalogue itself exhibits a flattening of the HOD at the high-mass end, with the number of satellites not increasing as expected. This deviates from the typical HOD behaviour, where the satellite population continues to grow with halo mass, following a power-law slope. Notably, the BST-recovered HOD reproduces this same trend, demonstrating its ability to capture the underlying distribution even in regimes affected by modelling.

\subsection{Luminosity-based group size}
\label{section:SS_2}

Reliable determination of group sizes in flux-limited samples is challenging because the number of detected members decreases with distance, introducing strong selection effects \citep{Yang2008}. We aim to estimate the characteristic radius of the host halo used in BST from the observed luminosity of the central galaxy of the group ($r_{\mathrm{c}} ^{\mathrm{lum}}$), to avoid relying on parameters from the simulation which we would not have access to in real data.

\begin{figure}
    \includegraphics[width=\columnwidth]{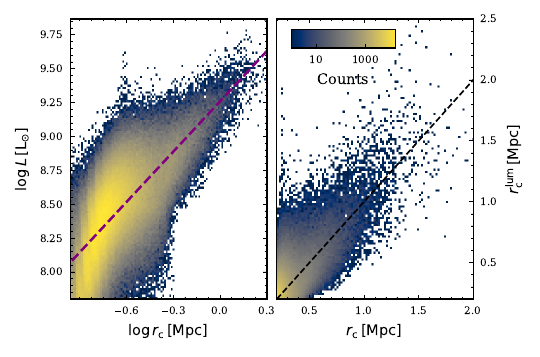}
    \caption{(Left panel) Relation between $r_{\mathrm{c}}$ and luminosity for central galaxies in our catalogue, with the best linear fit shown as purple dashed lines. (Right panel) Relation between $r_{\mathrm{c}}$ and the estimated $r_{\mathrm{c}}^{\mathrm{lum}}$, with the dashed line indicating the one-to-one relation.}
    \label{fig:fig8}
\end{figure}

We compute the linear fit between host halo characteristic radius ($r_{\mathrm{c}}$) and luminosity for central galaxies, as shown in the left panel of Fig.~\ref{fig:fig8}. This relation enables an interpolation scheme to infer $r_{\mathrm{c}}$, serving as a proxy for group size, directly from the total luminosity of a system. This approach is particularly useful when only photometric data are available.

We propose the relation 

\begin{equation}
    r_{\mathrm{c}}^{\mathrm{lum}} = C_{1} \cdot L + C_{2} 
\label{eq:2}
\end{equation}

\noindent where $C_{1} = 6.1 \times 10^{-10} \, \mathrm{Mpc}/\mathrm{L}_{\odot}$ and $C_{2} = -0.03 , \mathrm{Mpc}$. We show the comparison of $r_{\mathrm{c}}^{\mathrm{lum}}$ with $r_{\mathrm{c}}$ in the right panel of Fig.~\ref{fig:fig8}. Although both characteristic halo radii can be approximately related through a linear relation, the comparison reveals a significant dispersion which is expected to impact the HOD estimation.

\begin{figure}
    \centering    
    \includegraphics[width=\columnwidth]{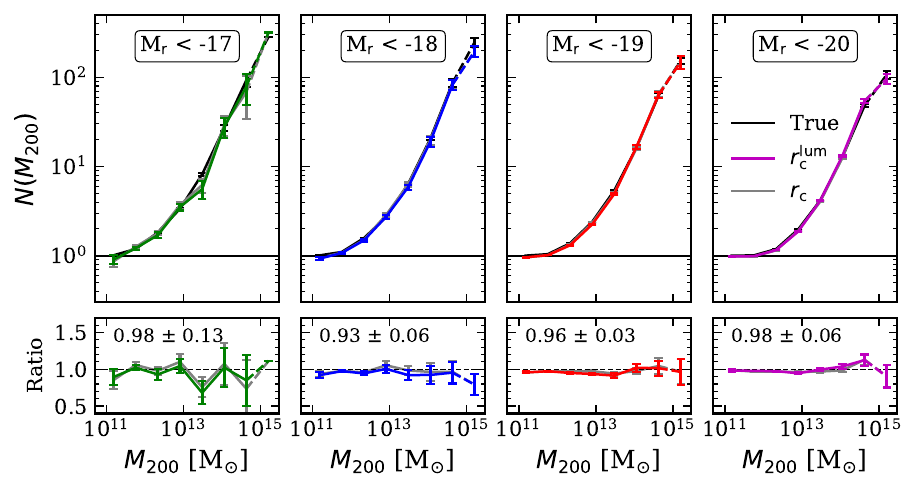}
    \caption{BST results using $r_{\mathrm{c}}^{\mathrm{lum}}$ derived from the central galaxy luminosity using Eq.~\ref{eq:2}. In gray lines, recovered HOD when using $r_{\mathrm{c}}$ and in black lines, the true HOD. The lower panels show the ratio between the BST-recovered and true HOD, together with the corresponding median and dispersion.}
    \label{fig:fig9}
\end{figure}

The results when using $r_{\mathrm{c}}^{\mathrm{lum}}$ instead of $r_{\mathrm{c}}$ show good agreement between the recovered HOD and the true HOD across the full halo mass range, presented in Fig.~\ref{fig:fig9}. While adopting a more flexible relation between central galaxy luminosity and size—such as a broken power law or a mass-dependent slope—could potentially yield even better results, our current linear parametrisation proves sufficiently accurate and robust for this analysis. This simple model is particularly valuable given the challenge of estimating the group radius, which typically requires a prior determination of its halo mass. In contrast, our approach allows a practical and effective estimation of group size directly from the luminosity of the central galaxy, enabling a reliable recovery of the HOD without detailed mass modelling. In the following sections, we therefore adopt $r_{\mathrm{c}}^{\mathrm{lum}}$, computed from the luminosity of the central galaxies, for the validation of the method.

\subsection{Photometric redshift}
\label{section:SS_3}

In this subsection, we investigate the impact of using photometric redshifts by adopting redshift estimates from FlexZBoost and BPZ. Specifically, we replace the \textit{true} redshift from the group centres, $z_{\mathrm{gr}}$, with the photometric redshift, $z_{\mathrm{photo}}$, and for estimating $r^{\mathrm{lum}}_{c}$ in BST through Eq.~\ref{eq:eq1}. This substitution introduces uncertainty in the absolute magnitude of central galaxies and their associated groups, significantly complicating the recovery of the HOD. Moreover, some migration of galaxies between the volume-limited subsamples is expected once photometric redshifts are applied, as will occur in real observational data.

\begin{figure}
    \centering    
    \includegraphics[width=\columnwidth]{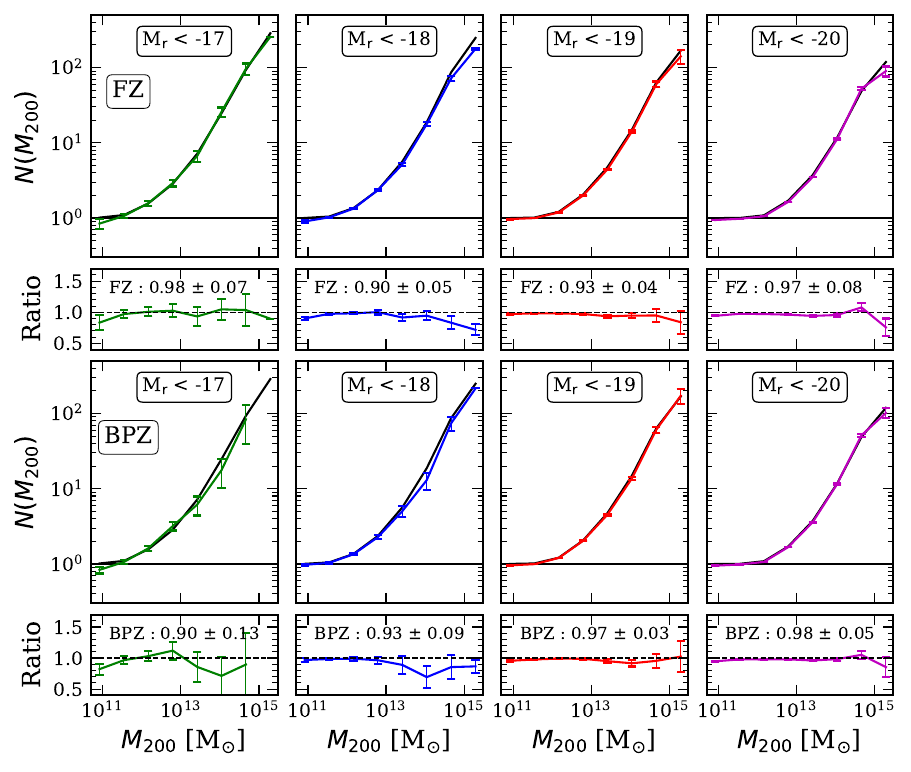}
    \caption{(Top panels)  Recovered BST HODs using FlexZBoost (FZ) photometric redshifts. The black lines show the true HOD. The lower panels show the ratio between the BST-recovered and true HOD, together with the corresponding median and dispersion. (Bottom panels) same as top but for BPZ.}
    \label{fig:fig10}
\end{figure}

In this test, we do not apply any quality cut on the photometric redshifts, allowing us to test the robustness of the BST with respect to photometric redshift in the ``worst" case scenario---i.e., with no mitigation of the impact of outliers on the final results. The resulting HODs obtained using the BST under these conditions are presented in Fig.~\ref{fig:fig10}. Both photometric redshift estimators are generally in good agreement with the truth across all magnitude cuts. BPZ yields lower performance than FlexZBoost, especially in the faintest cut ($M_{\mathrm{lim}}=-17$), due to more dispersed and biased photometric redshifts with respect to the true redshifts.

Uncertainties in the photometric redshifts of group centres do not contribute significantly to the background subtraction estimation of the total HOD. We note that, as shown in Fig.~\ref{fig:fig10}, the HOD inferred for groups identified using photometric redshifts exhibits only a mild underestimation at the high-mass end, while otherwise accurately reproducing the true HOD.

\subsection{Halo mass estimation}
\label{section:SS_4}

In photometric galaxy surveys, estimating halo masses is inherently challenging due to projection and miscentering effects, limited group membership, and uncertainties in galaxy properties. For instance, \citet{Yang2021} obtained halo mass estimates for each group by applying an abundance matching method using total group luminosity ranking, which links stellar mass to halo mass based on cumulative number density matching \citep[e.g.;][]{Moster2018,Behroozi2019}. Although this empirical approach provides a physically motivated alternative to group-based dynamical mass estimators, accurate stellar masses, star-formation rates, and luminosities are required to obtain reliable halo mass estimates, typically with a mean scatter of $\sigma \sim 0.20$ dex \citep{VanKempen2025}. 

Relying solely on apparent magnitudes of central-galaxy candidates (without confirmed group membership) substantially increases the uncertainties, making it challenging to robustly recover the HOD. An additional source of error arises from redshift uncertainties inherent in the use of photometric catalogues. To assess how such uncertainties might impact our method, we performed a controlled test in which we added a Gaussian dispersion of $0.1$ and $0.2$ dex to the simulated halo masses. These values reflect the typical maximum scatter observed in the empirical relation between the halo mass and the number of group members. Fig.~\ref{fig:fig_apx4} (Appendix~\ref{sec:apx_Mass}) shows the resulting redistribution of halo masses produced by these uncertainties. The associated uncertainty is largest at the low- and high-mass ends, and decreases towards the intermediate mass range.

\begin{figure}
    \centering    
    \includegraphics[width=\columnwidth]
    {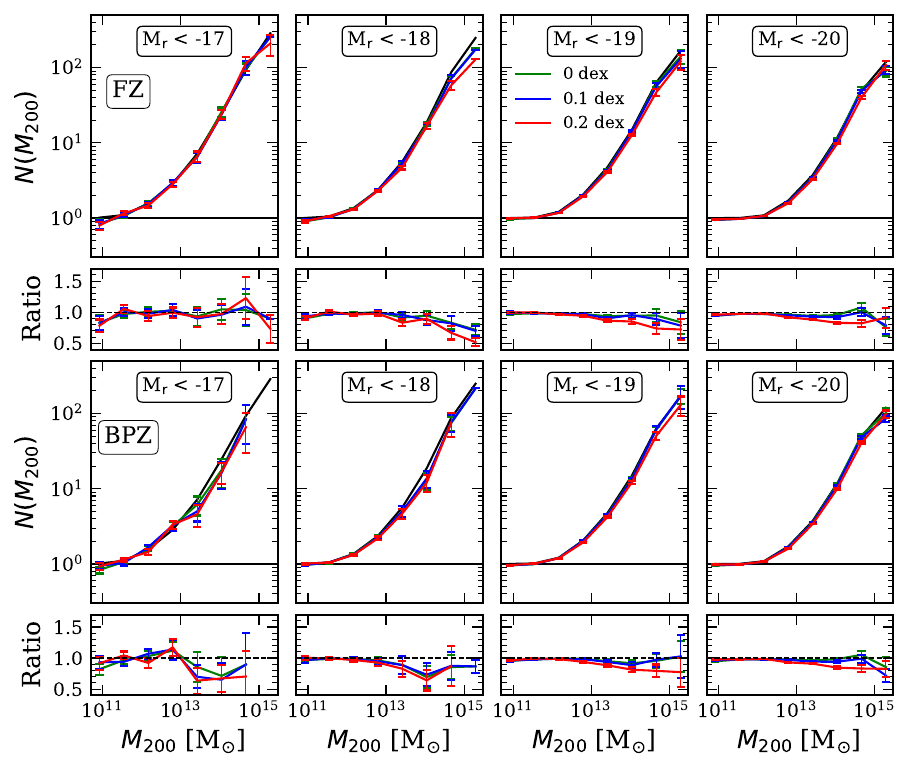}
    \caption{(Top panels) Recovered HODs using the BST under varying levels of Gaussian uncertainty added to the halo masses, expressed in dex units for FlexZBoost photometric redshifts. (Bottom panels) same as top but for BPZ. Increasing the scatter predominantly affects the high-mass end of the HOD, while the low-mass regime remains largely stable. }
    \label{fig:fig11}
\end{figure}

Figure~\ref{fig:fig11} presents the results of these tests. As shown, the added halo mass uncertainties do not produce a large impact on the recovered HOD, although small deviations become apparent at the high-mass end for the largest mass perturbation considered. BPZ performance weakens, particularly for the fainter magnitude cuts. The overall good agreement between the HODs obtained under different levels of halo mass scatter indicates that the method is robust against typical observational errors in halo mass estimation. 

\subsection{Application of the CGF}
\label{section:SS_5}

In this final subsection, we incorporate the CGF technique into the previously BST recovered HOD to evaluate the impact of group-centre misidentification in combination with all systematic effects considered so far (luminosity-based group size, photometric redshift errors, and a halo mass uncertainty of $0.1$ dex). We use CGF fixed parameters $(f_r,f_z) = (2.5,0.25)$, motivated by previous analysis (see Sect.~\ref{section:CGF}). Consequently, the total number of galaxies identified as central decreases significantly, primarily due to the lower \textit{completeness} associated with these conservative selection parameters. This reduction is quantified in Tab.~\ref{tab:tab2}, where the decrease in the number of selected central reflects the trade-off between \textit{completeness} and \textit{purity}. Nonetheless, the gain in \textit{purity} justifies this choice, as it enables a more reliable reconstruction of the true HOD. 

\begin{table}
\caption{Central galaxies identified with the set parameters $(f_r,f_z) = (2.5,0.25)$ from the CGF method. Each column represents a different selection based on magnitude limits and \textit{completeness} cuts. We also show identified central galaxies with Odds>0.8.}
\begin{center}
\small
\renewcommand{\arraystretch}{1.3}

\begin{tabularx}{\columnwidth}{X|X|X|X|X}
\hline
$M_{\mathrm{lim}}$ & $z_{max}$ & $N^{CGF}_{cen}$ & $N^{Odds}_{cen}$ \\
\hline
$-17$ & $0.088$&   $2.9 \cdot 10^{3}$ & $2.3 \cdot 10^{3}$\\
$-18$ & $0.135$ &  $6.8 \cdot 10^{3}$& $5.8 \cdot 10^{3}$\\
$-19$ & $0.206$ &  $1.3 \cdot 10^{4}$& $1.2 \cdot 10^{4}$\\
$-20$ & $0.304$ &  $2.0 \cdot 10^{4}$& $1.9 \cdot 10^{4}$\\
\hline
\end{tabularx}
\end{center}
\label{tab:tab2}
\end{table}

\begin{figure}
    \centering    
    \includegraphics[width=\columnwidth]{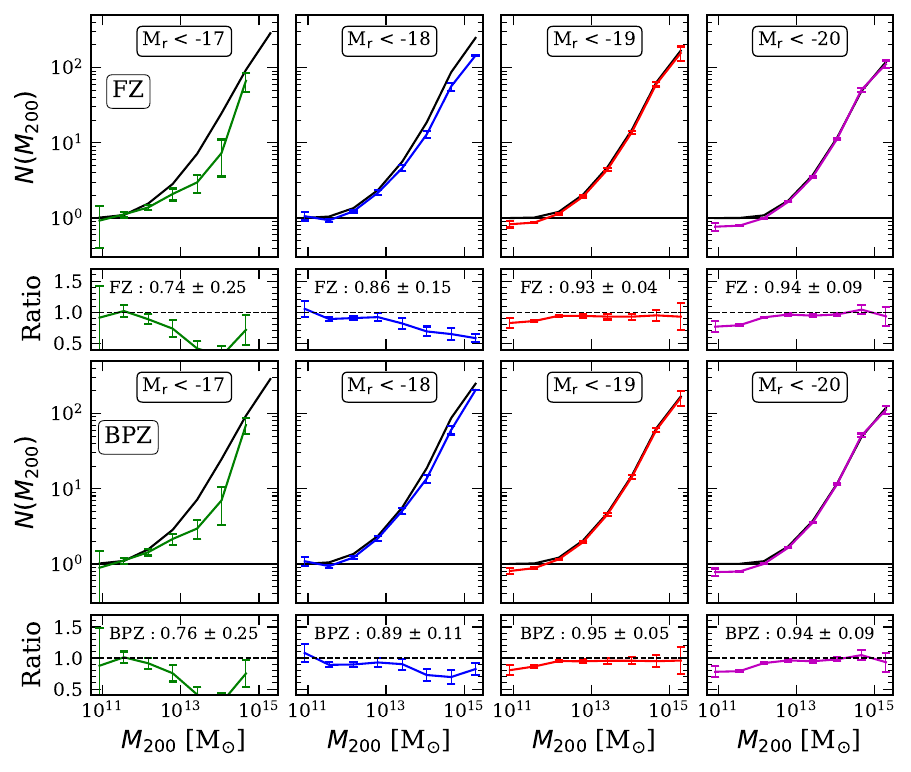}
    \caption{(Top panels) Results of the BST applied to galaxies identified with the CGF, with parameters $(f_r,f_z) = (2.5, 0.25)$ for FlexZBoost photometric redshifts. The black lines show the true HOD. (Bottom panels) same as top but for BPZ.}
    \label{fig:fig12}
\end{figure}

In Fig.~\ref{fig:fig12}, we present the results of applying the BST to this refined sample of central galaxies. We further restricted to objects with Odds>0.8, to reduce the contribution of photometric redshift outliers while preserving a high fraction of the original CGF-identified central galaxies. The HOD recovery shows good agreement with the true underlying distribution for the brighter magnitude limits ($M_{\mathrm{lim}} =-19$, $-20$). However, performance degrades for the other subsamples, which can be attributed to the low number of galaxies in this subsample (fewer than $10^4$). Due to reduced statistics from CGF application, achieving good agreement at the faint end remains challenging for FlexZBoost and BPZ, but both perform well in the brighter cuts. Larger simulations with broader sky coverage—such as a potential extension of \textsc{SkySim5000}—would provide improved sampling and enable more robust constraints on the HOD in this regime. Additionally, we note that more sophisticated group-centre identification techniques may yield further improvements; however, the approach adopted here provides satisfactory performance while maintaining a relatively simple methodology.

We further investigate the impact of varying the CGF parameters, $f_r$ and $f_z$, within the BST framework. Fig.~\ref{fig:fig_apx3} (Appendix~\ref{sec:apx_CGF}) shows the effect of changing these parameters on the recovered HOD. We find that increasing $f_r$ leads to better agreement between the recovered and true HOD, primarily through an improvement in the \textit{purity} of central galaxy identification. This effect is particularly relevant at high halo masses, where contamination from misidentified centrals can strongly bias the HOD estimate. In contrast, variations in the redshift discard parameter $f_z$ have a comparatively minor impact on the BST results, indicating that the method is less sensitive to this parameter within the explored range.

\section{Conclusions}
\label{section:conclusions}

\begin{table*}
\caption{Summary of the BST performance across all tests conducted in this work. Each column lists the characteristic halo radius used ($r_{\mathrm{c}}$), the method adopted to identify group centres, redshift employed for group centres, and the median ratio between the BST–recovered and true HOD values, $\left<\mathrm{HOD}_{\mathrm{BST}} / \mathrm{HOD}_{\mathrm{true}}\right>$. We only show results for 0.2 dex uncertainty in halo mass. The final column references the corresponding section where each test is discussed.}
\label{tab:tab3}
\begin{center}
\small
\renewcommand{\arraystretch}{1.3}
\begin{tabularx}{\textwidth}{
    l |
    p{1.cm} |
    p{2.1cm} |
    p{1.1cm} |
    c c c c |
    p{1.8cm}
}
\hline
 & Radius & Group Centres & Redshift &
 \multicolumn{4}{c|}{$\left< \mathrm{HOD}_{\mathrm{BST}}/\mathrm{HOD}_{\mathrm{true}} \right>$} & Sect. \\
\cline{5-8}
 &  &  &  &
 $M_r<-17$ & $M_r<-18$ & $M_r<-19$ & $M_r<-20$ &
   \\
\hline
Simulation-Based (Truth)
 & $r_{\mathrm{200}}$
 & True centrals
 & $z_{\mathrm{true}}$
 & $1.02 \pm 0.15$ & $0.97 \pm 0.07$ & $0.97 \pm 0.03$ & $0.97 \pm 0.06$
 & Sect.~\ref{section:SS_1} \\ [-0.8ex]

Luminosity-Based
 & $r^{\mathrm{lum}}_{\mathrm{c}}$
 & True centrals
 & $z_{\mathrm{true}}$
 & $0.98 \pm 0.13$  & $0.93 \pm 0.06$ & $0.96 \pm 0.03$ & $0.98 \pm 0.06$
 
 & Sect.~\ref{section:SS_2} \\

photo-$z$ (FlexZBoost)
 & $r^{\mathrm{lum}}_{\mathrm{c}}$
 & True centrals
 & $z_{\mathrm{photo}}$
 & $0.98 \pm 0.07$ & $0.90 \pm 0.05$  & $0.93 \pm 0.04$ & $0.97 \pm 0.08$
 & Sect.~\ref{section:SS_3} \\ [-0.8ex]
 
photo-$z$ (BPZ)
 & $r^{\mathrm{lum}}_{\mathrm{c}}$
 & True centrals
 & $z_{\mathrm{photo}}$
 & $0.90 \pm 0.13$ & $0.93 \pm 0.09$  & $0.97 \pm 0.03$ & $0.98 \pm 0.05$
 & Sect.~\ref{section:SS_3} \\ 

Error Mass (FlexZBoost)
 & $r^{\mathrm{lum}}_{\mathrm{c}}$
 & True centrals
 & $z_{\mathrm{photo}}$
 & $0.93 \pm 0.11$ & $0.93 \pm 0.21$  & $0.89 \pm 0.19$  & $0.90 \pm 0.08$
 & Sect.~\ref{section:SS_4} \\ [-0.8ex]
 
Error Mass (BPZ)
 & $r^{\mathrm{lum}}_{\mathrm{c}}$
 & True centrals
 & $z_{\mathrm{photo}}$
 & $0.93 \pm 0.19$ & $0.89 \pm 0.19$  & $0.94 \pm 0.08$  & $0.92 \pm 0.07$
 & Sect.~\ref{section:SS_4} \\

 CGF (FlexZBoost)
 & $r^{\mathrm{lum}}_{\mathrm{c}}$
 & CGF
 & $z_{\mathrm{photo}}$
 & $0.74 \pm 0.25$  & $0.86 \pm 0.15$  & $0.93 \pm 0.04$  & $0.94 \pm 0.09$
 & Sect.~\ref{section:SS_5} \\  [-0.8ex]
 
 CGF (BPZ)
 & $r^{\mathrm{lum}}_{\mathrm{c}}$
 & CGF
 & $z_{\mathrm{photo}}$
 & $0.76 \pm 0.25$  & $0.89 \pm 0.11$  & $0.95 \pm 0.05$  & $0.94 \pm 0.09$
 & Sect.~\ref{section:SS_5} \\ 
\hline
\end{tabularx}
\end{center}
\end{table*}

In this paper, we extend the Background Subtraction Technique (BST) developed by \citet{Rodriguez2015} to enable its application to galaxy samples based on both true and photometric redshifts. By incrementally introducing observational uncertainties into the variables involved in BST, we evaluated the ability of the method to recover the Halo Occupation Distribution (HOD). The main modifications and improvements to the original BST are summarised as follows:

\begin{itemize}
    \item We first evaluated the method by applying BST to the true values of redshift, halo mass, and radii in \textsc{cosmoDC2} and successfully recovered the HOD with excellent agreement. Following the classical BST approach, we used the true central galaxies as group centres with known positions.
    
    \item We then replaced the characteristic radius $r_{\mathrm{c}}$ with a more observationally motivated proxy, using the luminosity of the central galaxy ($r_{\mathrm{c}}^{\mathrm{lum}}$), showing that the luminosity-based scaling provides a sufficiently reliable proxy for group sizes within the context of our analysis.

    \item Next, we then added uncertainty to the group positions by incorporating photometric redshift errors. Despite a significant loss in statistical power due to the limited sky coverage of the DC2 simulation, we were still able to recover the HOD. When introducing typical uncertainties in the halo mass estimation, the HOD recovery remained robust.    
    
    \item We substituted the known central galaxies from the simulation with an inferred central-galaxy identification method (Central Galaxy Finder, CGF), based on galaxy luminosity and spatial distributions. We achieved a central-galaxy average \textit{purity} greater than 70 \%, with \textit{completeness} average over 60 \% over the halo mass range studied. This simple method effectively recovers the HOD for the brighter cuts, bringing our approach closer to observational techniques for constraining halo occupation.
    
\end{itemize}

Table~\ref{tab:tab3} summarises the BST performance across all tests. We recovered the HOD of the \textsc{cosmoDC2} catalogue in four ranges of limiting absolute magnitudes which extend from -17 to -20 in LSST r-band and halo masses up to $10^{15}\mathrm{M_{\odot}}$. 

These results demonstrate that the BST can be effectively extended to a broad range of observational datasets, including those with mixed redshift quality and limited group information. Our updated implementation reduces model dependence by relying on observable-based proxies and inference techniques, rather than requiring full spectroscopic group catalogues. This makes the method particularly well-suited for application to deep, wide-field photometric surveys such as LSST, Euclid, and Roman, which will lack complete spectroscopic coverage but offer unprecedented galaxy statistics.

This analysis is still far from the constraining power of an LSST-like survey: the increase in total area from $440$ to $18,000~\mathrm{deg}^2$ corresponds to an enhancement of the galaxy sample by a factor of $\sim 41$, and therefore to an improvement of $\sim \sqrt{41}$ in the statistical constraining power of HOD-based analyses. We note, however, that the galaxy samples used in this work are drawn directly from the \textsc{cosmoDC2} simulation, which represents an idealised dataset unaffected by atmospheric or instrumental effects. Although this work provides a thorough rundown of the major systematics relevant to our methodology, stars and blending are not modelled in \textsc{cosmoDC2}; nonetheless, our magnitude range ($m_r < 21$) lies well above the regime where blending significantly affects the detection of ultra-faint galaxies. The catalogue also does not include extinction from Milky Way dust or stellar foregrounds, modelling only the reddening due to dust within the host galaxies.

Importantly, the ability to recover the HOD from more photometry-based data opens the door to a wide range of science applications beyond traditional clustering analyses. By enabling the measurement of satellite occupation as a function of secondary galaxy properties—such as colour, local density, star formation rate, and morphology—this approach can contribute to constraining assembly bias and understanding the connection between galaxy evolution and environment. Furthermore, it provides a valuable framework for investigating phenomena such as galactic conformity, where the properties of satellite galaxies correlate with those of their central galaxy. We plan to explore these aspects further with the upcoming LSST data. Importantly, the present work also helps pave the way for the development of cluster-based analyses by providing a pipeline that is readily adaptable to LSST data releases.

\begin{acknowledgements}
PC wrote most of the manuscript and created and co-designed all figures, with guidance and input from the co-authors. VC developed most of the CGF method and provided feedback on the results and the manuscript. FR supervised the lead author and contributed to conceptualisation, methodology, project administration, resources, validation, and writing. AT developed the initial data selection, advised on project progress and presentation, and edited the manuscript. MCA edited the manuscript, provided substantial comments and suggestions during the writing process, contributed to the organisation of the paper, and advised on the presentation of the results. BL provided feedback as internal reviewer, leading to updates to manuscript structure and content. This paper has undergone internal review in the LSST Dark Energy Science Collaboration. We are very grateful to the internal reviewers: Heidi Wu and Benjamin Levine. PC acknowledges partial support from ANPCyT through grant PICT 2020-00582. MCA acknowledges support from ANID BASAL project FB210003. VC, FR and AT thank the support by Agencia Nacional de Promoci\'on Cient\'ifica y Tecnol\'ogica, the Consejo Nacional de Investigaciones Cient\'{\i}ficas y T\'ecnicas (CONICET, Argentina) and the Secretar\'{\i}a de Ciencia y Tecnolog\'{\i}a de la Universidad Nacional de C\'ordoba (SeCyT-UNC, Argentina). FR acknowledges support from the ICTP through the Junior Associates Programme 2023-2028. AT acknowledges support from the Dirección General de Asuntos del Personal Académico (DGAPA) of UNAM through the postdoctoral fellowship, and acknowledges support from the PAIRS project via the SONATA grant no.2023/51/D/ST9/02919. 

DESC acknowledges ongoing support from the IN2P3 (France), the STFC (United Kingdom), and the DOE and LSST Discovery Alliance (United States). DESC uses resources of the IN2P3 Computing Center (CC-IN2P3--Lyon/Villeurbanne - France) funded by the Centre National de la Recherche Scientifique; the National Energy Research Scientific Computing Center, a DOE Office of Science User Facility supported under Contract No.\ DE-AC02-05CH11231; STFC DiRAC HPC Facilities, funded by UK BEIS National  E-infrastructure capital grants; and the UK particle physics grid, supported by the GridPP Collaboration. This work was performed in part under DOE Contract DE-AC02-76SF00515. 

We thank the developers and maintainers of the following software used in this work: NumPy (van der Walt et al. 2011), SciPy (Jones et al. 2001), Matplotlib (Hunter 2007), Astropy (Astropy Collaboration et al. 2013) and Jupyter (Kluyver et al. 2016). 

\end{acknowledgements}

\bibliographystyle{aa} 
\bibliography{biblio} 

\appendix
\section{CGF}
\label{sec:apx_CGF}

\begin{figure}[H]
    \centering    
    \includegraphics[width=\columnwidth]{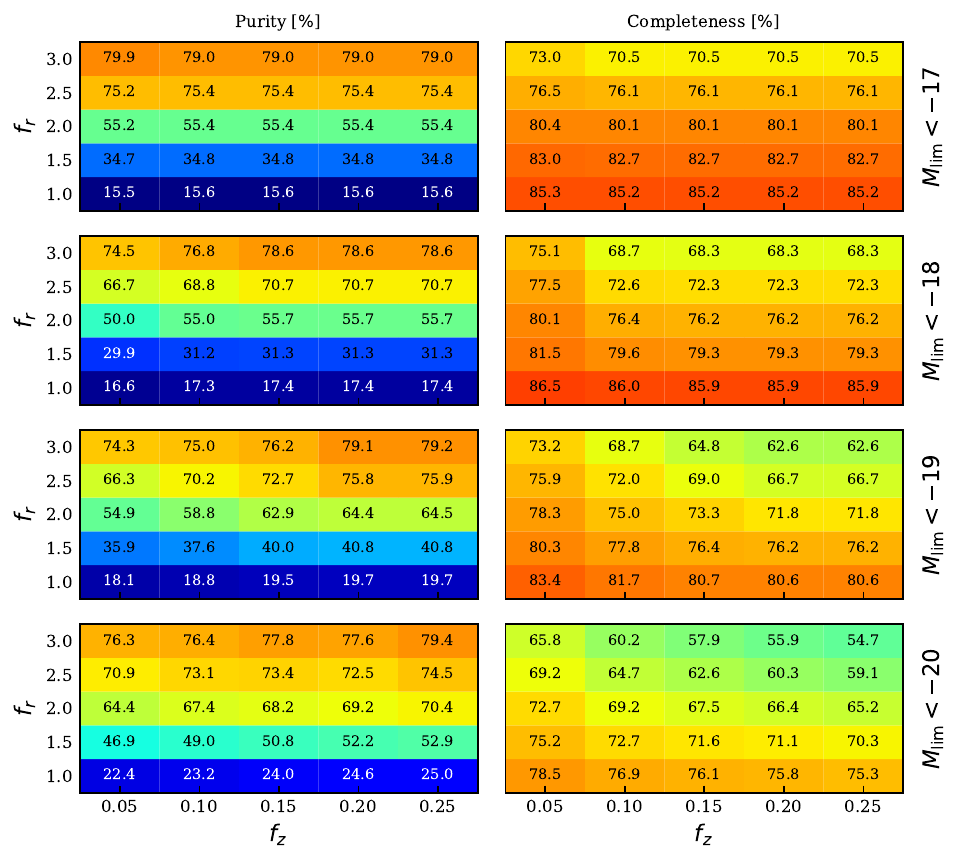}  
    \caption{Mean \textit{purity} (left panel) and \textit{completeness} (right panel) of the CGF, obtained by averaging these quantities over haloes with masses in the range $10^{12}$–$10^{15} \,\mathrm{M}_{\odot}$. Each panel corresponds to a different combination of the CGF parameters $f_r$ and $f_z$. Increasing $f_r$ and $f_z$ leads to higher mean \textit{purity} but lower mean \textit{completeness}, while smaller values improve \textit{completeness} at the cost of greater contamination.}
    \label{fig:fig_apx1}
\end{figure}

We explore the impact of varying the free parameters ($f_r,f_z$) in the CGF method. Fig.~\ref{fig:fig_apx1} shows the median \textit{purity} (top) and \textit{completeness} (bottom) of the CGF as a function of halo mass ($M_{\mathrm{200}}$) over the range $10^{12}$–$10^{15} \, \mathrm{M_{\odot}}$. The values in each cell of the table correspond to various combinations of the CGF parameters $f_r$ and $f_z$, which control the projected and redshift-space exclusion regions during iterative centre selection. 

As discussed in the main text, a clear trade-off emerges: higher values of $f_r$ and $f_z$ increase the exclusion volume, improving the purity—defined as the fraction of selected central candidates that are truly centrals—while reducing \textit{completeness}, the fraction of true centrals correctly identified. In contrast, lower parameter values yield better \textit{completeness} at the expense of \textit{purity}, as more contaminants are retained. 

\begin{figure}
    \centering    
    \includegraphics[width=\columnwidth]{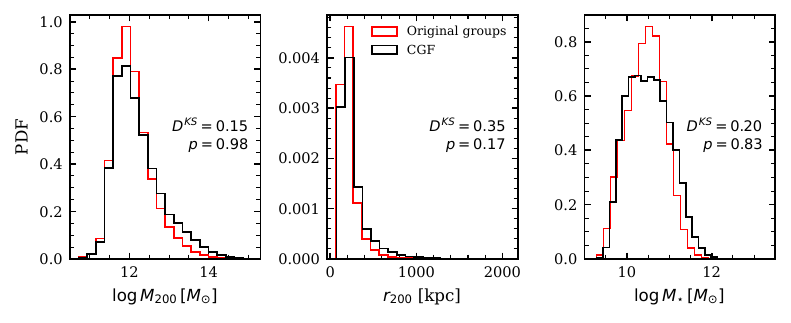} 
    \caption{Normalized distributions of halo mass $M_{\mathrm{200}}$ (left panel), halo radius $r_{\mathrm{200}}$ (middle panel), and stellar mass $M_{\star}$ (right panel) of CGF vs. the original groups sample. KS statistics and \textit{p}-values for each comparison are shown in the panels.}
    \label{fig:fig_apx2}
\end{figure}

These results illustrate the motivation behind our adopted parameter choice of $(f_r,f_z)=(2.5,0.25)$, which maintains \textit{purity} above $\sim$70 \% while preserving acceptable \textit{completeness}, especially important in the context of large upcoming photometric surveys such as LSST. This combination of parameters also yields a distribution of identified groups that is broadly consistent with those found using the original group sample, as shown in Fig.~\ref{fig:fig_apx2}. For this comparison, we compute the Kolmogorov–Smirnov \citep[KS,][]{Lederman1984} statistic ($D^{KS}$) over the full dataset to quantify the maximum deviation between the CGF-identified group centres and those in the original sample. The resulting \textit{p}-values ($p \gg 0.05$) indicate no significant difference between the two distributions, supporting the accuracy of the CGF in recovering the underlying group-centre distribution. However, it is important to note that this comes at the cost of a substantial reduction in the number of groups included in the CGF sample, reflecting the trade-off between \textit{purity} and completeness of the CGF method.

\begin{figure}
    \centering
    \includegraphics[width=\columnwidth]{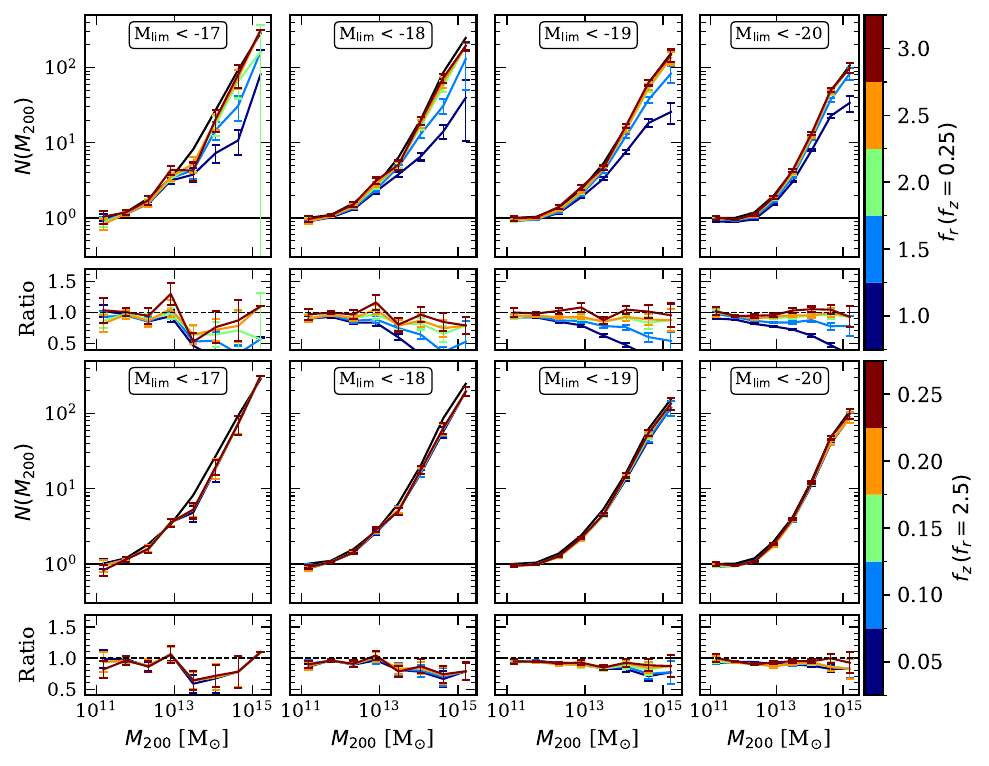}
    \caption{(Top panels) BST results using a fixed redshift parameter $f_z = 0.25$ and varying $f_r$ parameter. (Bottom panels) Results using a fixed $f_r = 2.5$ and varying the redshift parameter $f_z$. The performance shows better agreement with the true HOD as $f_r$ increases, while it does not change significantly when varying $f_z$.}
    \label{fig:fig_apx3}
\end{figure}

To further evaluate the impact of the CGF parameters on the accuracy of central galaxy identification, we conduct the BST varying one parameter at a time while fixing the other to a representative pivot value ($f_r = 2.5$ or $f_z = 0.25$). The results, shown in Fig.~\ref{fig:fig_apx3}, highlight how greater discard radii (specifically $r_{\mathrm{dsc}} < 2.5\, r_c^{\mathrm{lum}}$) lead to better agreement with the HOD. This improvement is directly related to the resulting \textit{purity} of the CGF, which is essential to faithfully recover the underlying HOD signal. 

Only for $f_r = 2.5$ and $f_r = 3.5$, where the purity exceeds roughly 70 \%, we obtain good agreement with the true HOD. This is particularly important in the high-mass regime, where the HOD is more sensitive to contamination. This analysis supports our choice of parameters for the main catalogue used in the following sections.

\section{Inference of the halo mass}
\label{sec:apx_Mass}

\begin{figure}[H]
    \centering
    \includegraphics[width=0.8\columnwidth]{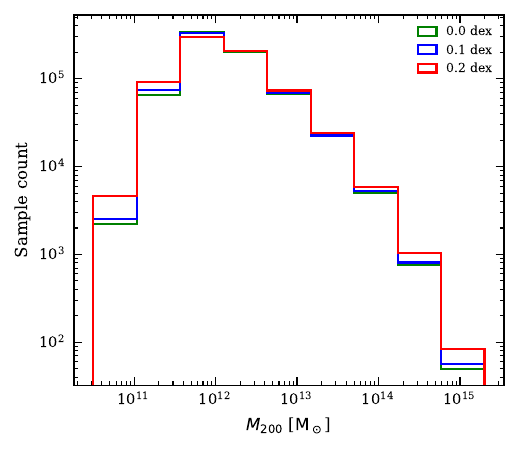}
    \caption{Distribution of halo masses in the group sample. The green curve shows the original distribution, while the blue and red curves represent the perturbed distributions obtained by adding Gaussian scatter in $M_{\mathrm{200}}$ with different standard deviations (in dex).}
    \label{fig:fig_apx4}
\end{figure}

To assess the impact of halo mass uncertainties on the overall distribution, we introduce Gaussian perturbations with varying standard deviations (in dex) to the original $M_{\mathrm{200}}$ values. As shown in Fig.~\ref{fig:fig_apx4}, increasing the level of scatter leads to a broader redistribution of halo masses. Notably, this results in a net increase in the number of haloes at both the low- and high-mass ends, as objects are scattered into the tails of the distribution from the central mass range.

\end{document}